\documentclass[12pt,twoside]{article}
\usepackage{amsmath,amsfonts,amssymb}
\usepackage{graphicx}
\usepackage{eucal}
\usepackage{mathrsfs}
\usepackage{theorem}
\usepackage{pifont}
\newcommand{\Real}{\mathbb{R}}

\newcommand{\todo}[1]{{\sffamily To do:}}

\newtheorem{theorem}{Theorem}
\newtheorem{proposition}{Proposition}

\newtheorem {corollary}{Corollary}
\newtheorem {lemma}{Lemma}
\newenvironment{proof}{{\flushleft \emph{Proof}:}}{\ding{110}}

\newenvironment{stasproof}{{\flushleft \emph{Proof of Theorem \ref{stas}}:}}{\ding{110}}
\newenvironment{proofw}{{\flushleft \emph{Proof of Theorem \ref{existence}}:}}{\ding{110}}
 \numberwithin{equation}{section} \numberwithin{theorem}{section}
\numberwithin{lemma}{section} \numberwithin{corollary}{section}
\numberwithin{proposition}{section}
\numberwithin{assumption}{section}

\date{}
\title{Synchronization of oscillators
coupled through an environment}
\author{Guy Katriel\footnotemark[1]}

\begin{document}

%
%
%

\maketitle
\renewcommand{\thefootnote}{\fnsymbol{footnote}}
\footnotetext[1]{Institute of Mathematics, The Hebrew University,
Jerusalem 91904, Israel.

Partially supported by the Edmund Landau Center for Research in
Mathematical Analysis and Related Areas, sponsored by the Minerva
Foundation (Germany).}

\begin{abstract}
We study synchronization of oscillators that are indirectly coupled
through their interaction with an environment. We give criteria for
the stability or instability of a synchronized oscillation. Using
these criteria we investigate synchronization of systems of
oscillators which are weakly coupled, in the sense that the
influence of the oscillators on the environment is weak. We prove
that arbitrarily weak coupling will synchronize the oscillators,
provided that this coupling is of the `right' sign. We illustrate
our general results by applications to a model of coupled GnRH
neuron oscillators proposed by Khadra and Li \cite{khadra}, and to
indirectly weakly-coupled $\lambda-\omega$ oscillators.
\end{abstract}




\section{Introduction}

The aim of this work is to investigate the dynamics of systems of
oscillators which are globally coupled through an
environment. An important example of such systems is populations of cells in which
oscillatory reactions are taking place \cite{golbeter,hess,kruse,schibler,winfree},
which `communicate' via chemicals that diffuse in
the surrounding medium. The ability of thousands of cells to synchronize their
periodic activity is crucial for the generation of macroscopic oscillations, like
circadian periodicities
\cite{foster}.

Consider a system of $n$ identical dynamical systems
(`oscillators'), described by the differential equations
\begin{equation}\label{un1}
\dot{x}_k=f(x_k,y),\;\;\;1\leq k\leq n,
\end{equation}
where $x_k\in \Real^d$ is the state of the $k$-th oscillator, $y\in
\Real^p$ is the state of the environment, and $f:\Real^d\times
\Real^p\rightarrow \Real^d$ is smooth. The dynamics of each
oscillator thus depends on the state of the environment. An
additional equation describes the dynamics of the environmental
variable $y$
\begin{equation}\label{un2}
\dot{y}=g(y)+\frac{\beta}{n}\sum_{j=1}^n h(x_j,y),
\end{equation}
where the smooth function $g:\Real^p\rightarrow \Real^p$ represents
the intrinsic dynamics of the environment, and the smooth function
$h:\Real^d\times \Real^p\rightarrow \Real^p$ represents the effect
of the oscillators on the environment. The state of the oscillators
thus influences the dynamics of the environment.

In the case of biological cells, $x_k$ would be a vector whose
components are the concentrations (in moles per unit volume) of
various biochemical species in cell $k$, and $y$ a vector of
concentrations of various biochemical species in the exterior of the
cells. The parameter $\beta$ is the ratio of the total intracellular
volume to the volume of the environment: if $V_{cell}$ is the volume
of an individual cell and $V_{ext}$ is the volume of the external
environment,
\begin{equation}\label{dbeta}
\beta=\frac{n V_{cell}}{V_{ext}}.
\end{equation}

In the case of biological cells, the interaction of cells and
environment may occur through the diffusion and transport of
chemical species across the cell membranes, and through the effects
of the activation of receptors on the cell membrane. A variety of
modeling studies of biochemical systems of oscillators coupled
through an environment, described by equations of the form
(\ref{un1}),(\ref{un2}), can be found in
\cite{camacho,furusawa,garcia,geier,gonze,henson,kuznetsov,madsen,toth,wang,wolf,wh,zhdanov}.
The framework presented above thus unifies many models of particular
systems, and allows us to obtain some basic analytical results which
apply to all of them.

Since the state of each of the oscillators influences the environment, and the state of the
environment in turn influences the oscillators, we obtain an {\it{indirect}} coupling
of the oscillators. We are interested in studying the capacity of this indirect coupling
to induce synchronization of the oscillators. When this happens, in the biochemical context,
one refers to the relevant species which diffuse in the environment as `synchronizing agents'.

By a {\bf{synchronized oscillation}} of the system (\ref{un1}),(\ref{un2})
we mean a periodic solution with all the $x_k$ identical
\begin{equation}\label{same}x_1(t)=x_2(t)=...=x_n(t)=x(t).
\end{equation}
The periodicity means that there exists a $T>0$ such that
\begin{equation}\label{per}
x(t+T)=x(t),\;\;\;y(t+T)=y(t),\;\;\;\forall t\in\Real.
\end{equation}
 Therefore, substituting (\ref{same}) into
(\ref{un1}),(\ref{un2}), we see that a synchronized oscillation
corresponds to a periodic solution of the system
\begin{equation}\label{sync1}
\dot{x}=f(x,y),
\end{equation}\begin{equation}\label{sync2}
\dot{y}=g(y)+\beta h(x,y).
\end{equation}
The original system (\ref{un1}),(\ref{un2}) is $nd+p$ dimensional,
whereas the system (\ref{sync1}),(\ref{sync2}) is only $d+p$
dimensional. Since (\ref{sync1}),(\ref{sync2}) do not depend on $n$,
a periodic solution of (\ref{sync1}),(\ref{sync2}) gives rise to a
synchronized oscillation of (\ref{un1}),(\ref{un2}) for {\it{any}}
$n\geq 1$. Note that system (\ref{sync1}),(\ref{sync2}) is simply
(\ref{un1}),(\ref{un2}) for the case $n=1$, so it describes the
behavior of a single oscillator placed in the environment, and we
can therefore refer to it as the {\bf{single-oscillator system}}.

In order for a synchronized oscillation to be observable, it must be
stable in the sense that it is asymptotically approached starting from an open set of
initial conditions (the precise definition is recalled in section
\ref{stability}). A crucial point must be made here: the
stability of a synchronized oscillation $x_1(t)=...=x_n(t)=x(t)$, $y(t)$
 refers to its stability as a
solution of the full system (\ref{un1}),(\ref{un2}), and does
{{\it{not}} follow from the stability of the corresponding solution
$(x(t),y(t))$ as a solution of the single-oscillator system
(\ref{sync1}),(\ref{sync2}). When we say that {\bf{synchronization}}
occurs, we mean that there exists a synchronized oscillation which
is {\it{stable as a solution of}} (\ref{un1}),(\ref{un2}). A
criterion for the (in)stability of a synchronized oscillation will
be derived in section \ref{stabc}.

\begin{figure}\label{ef}
\centering
    \includegraphics[height=14cm,width=6cm, angle=270]{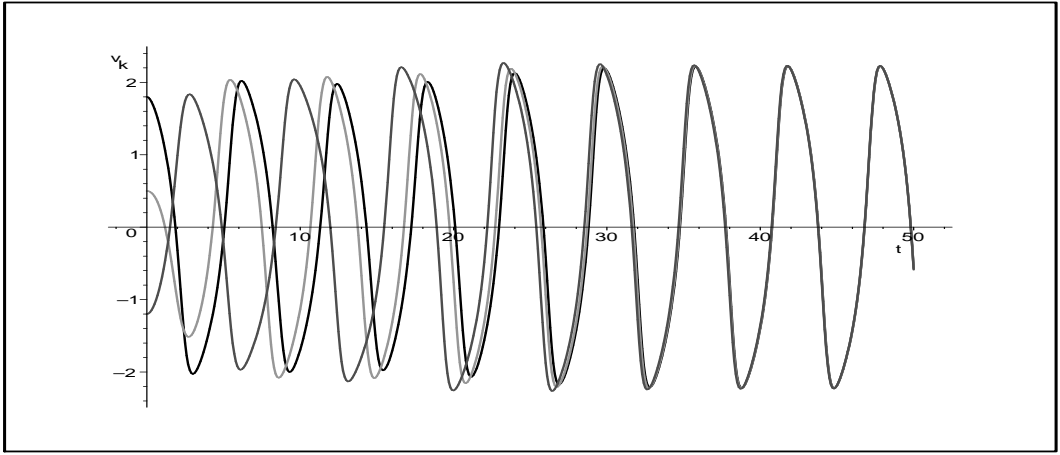}\\
    \vspace{1cm}
    \includegraphics[height=14cm,width=6cm, angle=270]{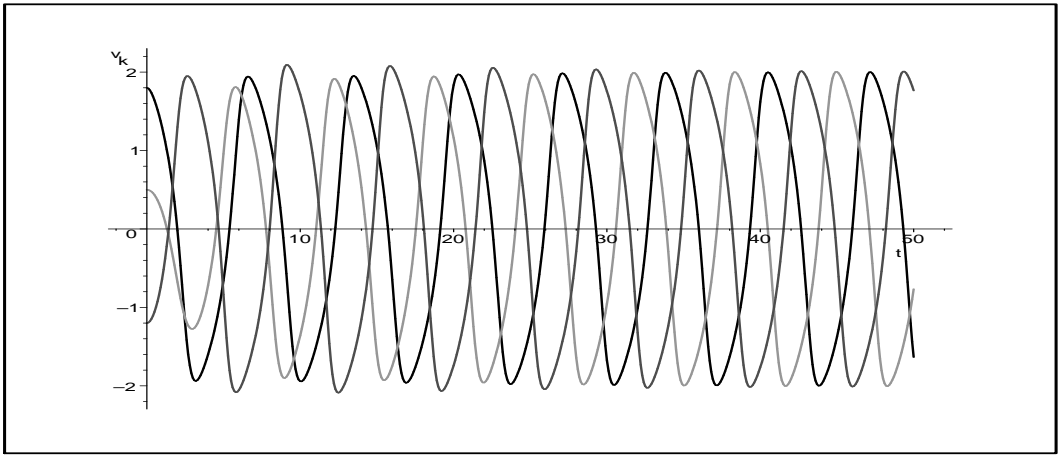}
    \caption{Three coupled Van der Pol Oscillators: Top:
    Graph of $v_k(t)$ vs. $t$ for $\beta=-0.5$. Bottom: $\beta=0.5$.
}
\end{figure}

In order to illustrate some of the collective phenomena observed in
systems of the form (\ref{un1}),(\ref{un2}) we display, in figure
\ref{ef}, the results of numerical simulations, in which we took
three Van der Pol oscillators coupled through an environment. The
dynamics of the oscillators are defined by
$$\ddot{v}_k+(v_k^2-1)\dot{v}_k+v_k=y,\;\;\;1\leq k\leq 3,$$
which we can convert to a first-order system
 (\ref{un1}), with $d=2$, $p=1$, $n=3$,
where $x=(v,w)$ and
$$f(x,y)=(w,(1-v^2)w-v+y).$$
We assume now that the environmental variable evolves according to (\ref{un2}), with
$$g(y)=-y,\;\;h(x,y)=v.$$
The system of the three oscillators coupled through the environment
is thus described by the equations
$$\ddot{v}_k+(v_k^2-1)\dot{v}_k+v_k = y,\;\;\;1\leq k\leq 3$$
$$\dot{y}=-y+\frac{\beta}{3}(v_1+v_2+v_3).$$
In the top part of figure \ref{ef} we set $\beta=-0.5$, and
plotted the values $v_k(t)$ ($1\leq k\leq 3$), starting at some
arbitrary initial conditions. One sees that the three oscillators
synchronize, so that within about 50 time units the three graphs
look identical.

On the other hand, in the bottom part of figure \ref{ef}, when we
set $\beta=0.5$ we see that the three oscillators do not
synchronize. In fact they seem to tend to the type of behavior known
as anti-synchronized or splay state, in which each of the three
oscillators perform the same motion, but with a phase lag of
$\frac{2\pi}{3}$ among the oscillators. When many oscillators are involved, such
behavior will lead to the averaging out of the oscillations, hence no
periodicity will be observable at the macroscopic level.

Our aim is to obtain some understanding of phenomena of
synchronization and desynchronization in oscillators coupled through
an environment, such as those demonstrated above. There is an
extensive literature concerned with the analysis of synchronization
of coupled oscillators (see \cite{hoppensteadt, manrubia,pikovsky}
and references therein), but most studies deal with directly coupled
oscillators, rather then oscillators indirectly coupled through an
environment. In some numerical and theoretical studies of systems of
the form (\ref{un1}),(\ref{un2}), `steady state' approximation on
(\ref{un2}) is made in order to transform it into a directly coupled
system: the term $\dot{y}$ on the left hand side of (\ref{un2}) is
replaced by $0$, the resulting algebraic equation is solved for $y$
in terms of $x$, and the expression for $y$ is substituted in
(\ref{un1}). Although in some cases this yields a good
approximation, it is not always so, and it is not hard to find
examples in which direct simulation shows synchronization of the
`approximate' system, while the system (\ref{un1}),(\ref{un2}) does
not synchronize, or vice versa. In this paper we study the
indirectly coupled system without this steady-state approximation.

The assumptions that the
dynamics of the $n$ oscillators, and their effects on the environment,
 are identical, should be considered
as idealizations which will only be satisfied in an approximate
sense in any real system. However, in studying phenomena such as
synchronization, it is useful to consider the heterogeneity of the
characteristics of the oscillators as a perturbation of an idealized
system of identical oscillators, and the results in this case are
robust in the sense that results for the idealized systems will also
explain and predict the behavior of heterogenous systems, at least
when the degree of heterogeneity is sufficiently small. Indeed, if
the system of identical oscillators has a stable synchronized
oscillation, then by the basic perturbation theory of periodic
solutions, any sufficiently close heterogeneous system, that is with
$f$,$g$,$h$ in (\ref{un1})-(\ref{un2}) replaced by
$f_k$,$g_k$,$h_k$, will have a stable periodic solution which is
`almost synchronized' in the sense that $|x_j(t)-x_k(t)|$ is small
for all $j,k$.

Another approximation implicit in the model (\ref{un1}),(\ref{un2})
is that the environment is homogeneous, which, in the biochemical
context, means that the various chemicals diffuse in the environment
on a time scale which is faster than that of the reactions in the
cells and the diffusion across the cell membranes, or alternatively
that the medium surrounding the cells is stirred.

In section \ref{stabc} we prove a basic result, Theorem \ref{stas}, which allows
to determine the (in)stability of synchronized oscillations of (\ref{un1}),(\ref{un2}) in
terms of the stability of two associated linear systems with periodic coefficients.
Notably, these two linear systems, hence
the stability of synchronized oscillations,
do not depend on the number $n$ of oscillators. Theorem \ref{stas} thus reduces arbitrarily
large problems to a pair of problems which are of fixed size, by exploiting symmetry.

In section \ref{weak} we make a general study of systems of oscillators which are {\it{weakly}}
coupled in the sense that they are described by (\ref{un1}),(\ref{un2}) with $|\beta|$ small.
Using Theorem \ref{existence} and some perturbation calculations, we prove that
a system of oscillators coupled through an environment can be synchronized using
an arbitrarily weak coupling, {\it{provided}} that $\beta$ is chosen to be of the `right' sign,
and derive a formula which tells us what this right sign is.

In sections \ref{gonadotropin} and \ref{hopf} we present two
examples of applications of our analytical results to specific
systems.

In section \ref{gonadotropin} we apply the general Theorem
\ref{stas} to the study of a model of periodic release of GnRH,
proposed by Khadra and Li \cite{khadra}. We show that whenever this
model, for a single cell, produces oscillations, the oscillations of
any number of cells coupled through the environment will
synchronize.

In section \ref{hopf} we apply Theorem \ref{existence}, which deals
with the weakly-coupled case, to the particular example of
indirectly coupled $\lambda-\omega$ oscillators, to derive explicit
conditions for (de)synchronization in the weak coupling regime.

In section \ref{discussion} we conclude with some comments on the
applicability of the results obtained in this paper to the study of
systems of the form (\ref{un1})-(\ref{un2}).

\section{Stability, instability and linearization}
\label{stability}

We recall the definitions of relevant notions of
stability. Let
\begin{equation}\label{general} \dot{z}=\Phi(z),
\end{equation}
where $\Phi:\Real^N \rightarrow \Real^N$, be a system of differential
equations. Let $\bar{z}(t)$ be a $T$-periodic solution of
(\ref{general}). We denote by $O\subset \Real^N$ the corresponding orbit
$$O=\{ \bar{z}(t)\;|\; t\in\Real\}.$$
$\bar{z}(t)$ is said to be {\bf{orbitally asymptotically stable}} if
there exists an open set $O\subset U\subset \Real^N$, so that for
any $z_0\in U$, there exists a $\rho$ such that the solution $z(t)$
of the initial value problem (\ref{general}), $z(0)=z_0$, satisfies
$$\lim_{t\rightarrow \infty}|z(t)-\bar{z}(t+\rho)|=0.$$

$\bar{z}(t)$ is said to be {\bf{unstable}} if there exists an open
set $O\subset U\subset \Real^N$ such that for any open set $O\subset
V\subset \Real^N$ there exists $z_0\in V$ and $\tilde{t}>0$ so that
the solution $z(t)$ of the initial value problem (\ref{general}),
$z(0)=z_0$, satisfies $z(\tilde{t})\not\in U$.

We now recall the notion of {\it{linearized stability}}. With a
periodic solution $\bar{z}(t)$ of (\ref{general}) we associate the
$T$-periodic linearized equation
\begin{equation}
\label{linearized} \dot{w}=\Phi'(\bar{z}(t))w.
\end{equation}
A Floquet multiplier of (\ref{linearized}) is a (generally complex)
number $\mu$ for which (\ref{linearized}) has a solution $w(t)$ satisfying
$$w(t+T)=\mu w(t).$$
In other words, defining the fundamental solution of
(\ref{linearized}) as the $N\times N$-matrix valued function $H(t)$
satisfying
$$\dot{H}(t)=\Phi'(\bar{z}(t))H(t),\;\;H(0)=I,$$
the eigenvalues of $H(T)$ are the Floquet multipliers associated
with (\ref{linearized}).

By differentiating (\ref{general}) with respect to $t$, we get
$$\ddot{\bar{z}}=\Phi'(\bar{z}(t))\dot{\bar{z}},$$
so that $\dot{\bar{z}}$ is a $T$-periodic solution of
(\ref{linearized}), which means that $\mu=1$ is always a Floquet
multiplier of the linearized equation. The periodic solution
$\bar{z}(t)$ is said to be {\bf{non-degenerate}} if $\mu=1$ is a
{\it{simple}} Floquet multiplier (that is, it is a simple eigenvalue
of the matrix $H(T)$). It is said to be {\bf{linearly stable}} if it
is non-degenerate and all Floquet multipliers other than $\mu=1$
have absolute values strictly less than $1$. It will be said to be
{\bf{linearly unstable}} if there is a Floquet multiplier with
absolute value strictly greater than $1$.

A fundamental result (see {\it{e.g.}} \cite{robinson}, Ch. V,
Theorem 8.4) states that
\begin{lemma}
(i) A linearly stable periodic solution is orbitally asymptotically
stable.

\noindent
(ii) A linearly unstable solution is unstable.
\end{lemma}
From now on we will refer to linearly (un)stable periodic solutions
simply as {\it{(un)stable}}.

It should be noted that the notions of stability defined above, and the
results which will be obtained below, are local. Stability of the synchronized oscillation
does not imply that synchronization will be reached from all initial conditions, but only
that it will be reached with some positive probability - that is for a set of initial conditions of
positive measure.

\section{Criterion for stability of a synchronized oscillation}
\label{stabc}

To study stability of a synchronized oscillation
$x_1(t)=\cdots=x_n(t)=x(t),y(t)$ of
\begin{equation}\label{s1}
\dot{x}_k=f(x_k,y),\;\;\;1\leq k\leq n,
\end{equation}\begin{equation}\label{s2}
\dot{y}=g(y)+\frac{\beta}{n}\sum_{j=1}^n h(x_j,y),
\end{equation}
where $x(t),y(t)$ is a $T$-periodic solution of the
single-oscillator system (\ref{sync1}),(\ref{sync2}),
 we
linearize the system (\ref{s1}), (\ref{s2}) around this solution, obtaining
\begin{equation}\label{ll1}
\dot{w}_k=f_x(x(t),y(t))w_k+f_y(x(t),y(t))z,\;\;\;1\leq k\leq n,
\end{equation}
\begin{equation}\label{ll2}
\dot{z}=\frac{\beta}{n}h_x(x(t),y(t))\sum_{j=1}^n  w_j +
[g'(y(t))+\beta h_y(x(t),y(t))]z.
\end{equation}

Although the system (\ref{ll1}),(\ref{ll2}) is an
$nd+p$-dimensional one, we will show below, using some simple linear
algebra, and exploiting the symmetry of the system with respect to permutation of the oscillators,
that the study of
its stability reduces to the study of the stability of two linear
systems, of dimensions $d+p$ and $d$ respectively. When $n$ is large
this is a huge reduction in the complexity of the problem.

\begin{theorem}\label{stas}
Let $x_1(t)=\cdots=x_n(t)=x(t),y(t)$ be a $T$-periodic synchronized
oscillation
 of (\ref{s1}),(\ref{s2}). This solution is stable if the
 following two conditions hold:
\begin{enumerate}
\item[(C1)] All of the $d$ Floquet multipliers of the $T$-periodic linear equation
\begin{equation}\label{pse0}
\dot{w}=f_x(x(t),y(t))w
\end{equation}
have absolute values less than $1$.

\item[(C2)]  The Floquet multiplier $\mu=1$ of the $T$-periodic linear
system
\begin{eqnarray}\label{eq2}
\dot{w}&=&f_x(x(t),y(t))w+f_y(x(t),y(t))z, \nonumber\\
\dot{z}&=&\beta h_x(x(t),y(t))w+[g'(y(t))+\beta h_y(x(t),y(t))]z
\end{eqnarray}
is simple, and all the other $d+p-1$ Floquet multipliers have
absolute values less than $1$.
\end{enumerate}
If one of the Floquet multipliers of either (\ref{pse0}) or
(\ref{eq2}) has absolute value greater than $1$, then the
synchronized oscillation is unstable.
\end{theorem}

There is an illuminating interpretation of the conditions (C1),(C2).
Condition (C2) says that $(x(t),y(t))$ is stable as a solution of
single-oscillator system (\ref{sync1}),(\ref{sync2}). As we noted in
the introduction, this is a much weaker condition then the statement
that the synchronized oscillation is stable as a solution of the
system (\ref{s1}),(\ref{s2}). However Theorem \ref{stas} tells us
that the only condition that we have to add in order to get this
stronger conclusion is (C1), that is the stability of the linear
system (\ref{pse0}). This is the system one would obtain by looking
at $x(t)$ as a periodic solution of the {\it{periodically forced}}
system
$$\dot{x}=f(x,y(t)),$$
with $y(t)$ considered as a given forcing, and asking for the
stability of $x(t)$ as a solution of this forced system.

Note that the conditions (C1),(C2) do not depend on $n$, so that we
see that stability of the synchronized solution $x_1=...=x_n=x$,
$y$, where $x$,$y$ is an oscillation of the single-oscillator system
(\ref{sync1}),(\ref{sync2}) does {\it{not}} depend on $n\geq 2$. In
other words, if two oscillators synchronize then any number of
oscillators will synchronize, provided the ratio $\beta$ of the
total intracellular volume to the volume of the environment is
maintained fixed.

In order to apply Theorem \ref{stas} to a particular system, one
needs to verify the conditions (C1) and (C2), so one needs to study
the nonautonomous periodic systems (\ref{pse0}) and (\ref{eq2}), a
task which may not be easy. Moreover, in general one does not have
an explicit expression for the periodic oscillation $\bar{x}(t)$, so
that even writing down these systems is not possible. Therefore in
general the verification of these conditions will involve numerical
computations. There are, however, systems for which the conditions
can be verified based on general considerations. Such an example is
presented in section \ref{gonadotropin}. In addition, Theorem
\ref{stas} is useful for deriving other general results, as we
demonstrate in the investigation of weakly coupled oscillators in
section \ref{weak}.

\begin{stasproof}
Written in matrix notation, the system (\ref{ll1}),(\ref{ll2}) is
\begin{equation}\label{lm1}
\begin{pmatrix}
  \dot{w}_1 \\
  \vdots \\
  \dot{w}_n \\
  \dot{z}
\end{pmatrix}
= A(t)
\begin{pmatrix}
  w_1 \\
  \vdots \\
  w_n \\
  z
\end{pmatrix},
\end{equation}
where
$$
A(t)=
\begin{pmatrix}
  a(t) & 0 & \cdots & 0 & b(t) \\
  0 & a(t) & 0 & 0 & b(t) \\
  0 & \cdots & \ddots & 0 & \vdots \\
  0 & \cdots & 0  & a(t) & b(t) \\
  c(t) & c(t) & \cdots & c(t) &  d(t)
\end{pmatrix},
$$
$$a(t)=f_x(x(t),y(t)),\;\;\;b(t)=f_y(x(t),y(t)),$$
$$c(t)=\frac{\beta}{n}h_x(x(t),y(t)),\;\;\;d(t)=g'(y(t))+\beta h_y(x(t),y(t)).$$

In order to determine the stability of the synchronized oscillation,
we wish to find the Floquet multipliers of (\ref{lm1}).

The solution of (\ref{lm1}) given by
\begin{equation}\label{fl0}
\begin{pmatrix}
  w_1 \\
  \vdots \\
  w_n \\
  z
\end{pmatrix}=
\begin{pmatrix}
  \dot{x} \\
  \vdots \\
  \dot{x} \\
  \dot{y}
\end{pmatrix},
\end{equation}
corresponds to the Floquet multiplier $1$. The synchronized
oscillation is linearly stable if this is the only solution
corresponding to the Floquet multiplier $1$, and the other Floquet
multipliers have absolute values less than $1$. We now display more
solutions of (\ref{lm1}) and their corresponding Floquet multipliers.
Assume that $w:\Real\rightarrow \Real^d$ satisfies
(\ref{pse0}). Then by direct inspection one sees that
\begin{equation}\label{sol1}\left(%
\begin{array}{c}
  w \\
  -w \\
  0 \\
  \vdots\\
  \vdots\\
  0 \\
\end{array}%
\right), \;
\left(%
\begin{array}{c}
  w \\
  0 \\
  -w \\
  0\\
  \vdots\\
  0 \\
\end{array}%
\right),\;\cdots\;,\;
\left(%
\begin{array}{c}
  w \\
  0 \\
  \vdots \\
  0\\
  -w\\
  0 \\
\end{array}%
\right)\end{equation} are linearly independent solutions of
(\ref{lm1}). Thus the $d$ Floquet multipliers of (\ref{pse0}) are
also Floquet multipliers of (\ref{lm1}), and condition (C1) implies
that these Floquet multipliers have absolute values less than $1$.
Since each of these Floquet multipliers correspond to $n-1$
eigenvectors, we have accounted for $(n-1)d$ of the $nd+p$ Floquet
multipliers.

We note also that if $w,z$ is a solution of (\ref{eq2}) then
$$
\begin{pmatrix}
  w \\
  \vdots \\
  w \\
  z
\end{pmatrix}
$$
is a solution of (\ref{lm1}). Thus the Floquet multipliers of
(\ref{eq2}) are also Floquet multipliers of (\ref{lm1}), and
condition (C2) implies that these Floquet multipliers, except the
one corresponding to (\ref{fl0}), have absolute values less than
$1$. Since (\ref{eq2}) is a $d+p$ system we have now another $d+p$
Floquet multipliers. We have thus accounted for all $nd+p$ Floquet
multipliers of (\ref{lm1}), and shown that under conditions (C1) and
(C2), the system (\ref{lm1}) is stable. The argument for instability
is similar.
\end{stasproof}

\section{Synchronization of oscillators weakly coupled through an environment}
\label{weak}

In this section we consider the system (\ref{s1}),(\ref{s2}), under the assumption that
the coupling of the oscillators to the environment is weak, in the sense that $|\beta|$ is
small. We will use Theorem \ref{stas}, and perturbation calculations, to
obtain information on the (in)stability of synchronized oscillations in this regime.

Note that, in the biochemical context, in view of (\ref{dbeta}) the
weak coupling case arises when the volume of the environment is
large relative to the total intracellular volume, hence the
secretion of a synchronizing agent into the environment by a cell
has only a weak effect on the concentration of this synchronizing
agent in the environment.

When $\beta=0$, the oscillators do not influence the
environment, hence they are also uncoupled from each other. We assume that in such a case
the environment has a steady state $\bar{y}$, and that when the
environment is in the state $\bar{y}$ the oscillators have
a periodic solution $\bar{x}(t)$:

\begin{assumption}\label{assume1}
\begin{enumerate}
\item[(i)] The equation
\begin{equation}\label{b01}\dot{y}=g(y)\end{equation}
has a stable steady state $\bar{y}$, that is $g(\bar{y})=0,$ and
that the eigenvalues of the matrix $$A=g'(\bar{y})$$ have negative
real parts.

\item[(ii)] The equation
\begin{equation}\label{b02}\dot{x}=f(x,\bar{y})\end{equation}
has a non-degenerate $T_0$-periodic solution $\bar{x}(t)$.
\end{enumerate}
\end{assumption}

Part (i) of Assumption \ref{assume1}, holds, for example, in the simplest case,
in which the dynamics in the environment is `trivial', consisting simply of the
decay of the various species, so that $g(y)=-Dy$, where $D$ is a diagonal
matrix whose diagonal coefficients are the various rates of decay, and $\bar{y}=0$.
We mention here a paper of Watanabe (see \cite{watanabe1}, Theorem 3),
which includes a perturbation result for synchronized
oscillations of systems of the form (\ref{s1}),(\ref{s2}) when $|\beta|\rightarrow 0$ in the
case that $g\equiv 0$, a case which is excluded by Assumption \ref{assume1}.

When $\beta=0$ the oscillators are uncoupled, so they will not be
able to synchronize, and the most we can expect, in case that
$\bar{x}(t)$ is a stable periodic solution of (\ref{b02}), is that
oscillator $k$ will tend
 to $\bar{x}(t+\rho_k)$ with different and unrelated values of $\rho_k$.
We shall show that for $|\beta|>0$ sufficiently small there is a synchronized oscillation,
but the stability of this oscillation depends on the sign of $\beta$: there exists a
number $\sigma$ such the synchronized oscillation is stable if $\sigma \beta>0$, and is unstable
if $\sigma \beta <0$. We shall give an explicit formula for computing $\sigma$.

\begin{theorem}\label{existence}
Assume that Assumption \ref{assume1} holds. Then there exists $\beta_0>0$ and smooth
functions $T(\beta)$, $x(\beta,t)$, $y(\beta,t)$, with
$$T: (-\beta_0,\beta_0)\rightarrow (0,\infty),$$
$$x: (-\beta_0,\beta_0)\times \Real \rightarrow \Real^d,\;\;\;y: (-\beta_0,\beta_0)\times \Real \rightarrow \Real^p,$$
and
$$T(0)=T_0,\;\;\;x(0,t)=\bar{x}(t),\;\;\;y(0,t)=\bar{y},$$
so that:

\noindent
(i) For all $|\beta|< \beta_0$, \begin{equation}\label{star}
x_1(t)=x_2(t)=...=x_n(t)=x(\beta,t),\;\;\;y(t)=y(\beta,t)\end{equation}
is a synchronized oscillation of the system
(\ref{s1}),(\ref{s2}), with period $T(\beta)$.

\noindent (ii) Letting $q(t)$ denote the $T_0$-periodic solution of
the linear equation
\begin{equation}\label{defq}
\dot{q}(t)=-[f_{x}(\bar{x}(t),\bar{y})]^*q(t),
\end{equation}
normalized so that
\begin{equation}\label{norm}
\int_0^{T_0} \langle q(s),\dot{\bar{x}}(s)\rangle ds=1,
\end{equation}
we have the following asymptotic formula for the period $T(\beta)$
of the above synchronized oscillation as $\beta\rightarrow 0$:
\begin{small}\begin{eqnarray}\label{pers}\frac{T(\beta)}{T_0} &=&1- \beta
\int_{0}^{\infty}\int_0^{T_0} \langle
f_y(\bar{x}(s),\bar{y})e^{rA}h(\bar{x}(s-r),\bar{y}), q(t)\rangle ds
dr+O(\beta^2).\nonumber\\
\end{eqnarray}\end{small}
\noindent (iii) If $\bar{x}(t)$ is stable as a periodic solution of
(\ref{b02}), then defining
\begin{equation}\label{dsig}\sigma= \int_0^{T_0}
\int_{0}^{\infty}\langle
f_y(\bar{x}(s),\bar{y})e^{rA}h_x(\bar{x}(s-r),
\bar{y})\dot{\bar{x}}(s-r),q(s)\rangle dr ds,\end{equation}
 we have
that, for $0<|\beta|<\beta_0$, (\ref{star}) is stable as a solution
of (\ref{s1}),(\ref{s2}) if $\sigma\beta>0$, and unstable if
$\sigma\beta<0$.

\noindent (iv) If $\bar{x}(t)$ unstable as a periodic solution of
(\ref{b02}), then (\ref{star}) is unstable as a solution of
(\ref{s1}),(\ref{s2}) for any $0<|\beta|<\beta_0$.
\end{theorem}

We recall a fundamental result regarding the perturbation of a
non-degenerate periodic solution of a differential equation, when the equation is perturbed.
\begin{lemma}\label{pertn}
Let
\begin{equation}\label{parameterized}
\dot{x}=\Phi(\beta,x),
\end{equation}
where $\Phi:\Real\times\Real^N\rightarrow \Real^N$ is smooth, be a
differential equation parameterized by the parameter $\beta$. Assume
that when $\beta=0$, (\ref{parameterized}) has a $T_0$-periodic
solution $\bar{x}(t)$, which is non-degenerate. Then there exists
$\beta_0>0$ such that if $|\beta|<\beta_0$ then
(\ref{parameterized}) has a non-degenerate $T(\beta)$-periodic
solution $x(\beta,\cdot)$, where $T:(-\beta_0,\beta_0)\rightarrow
(0,\infty)$ and $x:(-\beta_0,\beta_0)\times \Real^N\rightarrow
\Real^N$ are smooth functions satisfying
$$T(0)=T_0,\;\;\;x(0,t)=\bar{x}(t),\;\;\;\forall
t\in\Real,$$ and if $\bar{x}(t)$ is a linearly (un)stable solution
of (\ref{parameterized}) for $\beta=0$, then $x(\beta,\cdot)$ is a
linearly un(stable) solution of (\ref{parameterized}) for
$|\beta|<\beta_0$.
\end{lemma}

\begin{proofw}
The existence of the synchronized oscillation, that is a periodic solution of
\begin{equation}\label{sync1sm}
\dot{x}=f(x,y),
\end{equation}\begin{equation}\label{sync2sm}
\dot{y}=g(y)+\beta h(x,y)
\end{equation}
for $|\beta|$ sufficiently small, is an application of Lemma \ref{pertn}. Indeed, by Assumption \ref{assume1},
when $\beta=0$, (\ref{sync1sm}),(\ref{sync2sm})
has the non-degenerate $T_0$-periodic solution $x(t)=\bar{x}(t),\;y(t)=\bar{y}$, which is perturbed to
a $T(\beta)$-periodic solution $x(\beta,t)$,$y(\beta,t)$ of (\ref{sync1sm}),(\ref{sync2sm}) for $|\beta|$ small,
which gives the synchronized oscillation of (\ref{s1}),(\ref{s2}) and proves part (i).

If $\bar{x}(t)$ is stable as a periodic solution of (\ref{b02}), and
since all the eigenvalues of $A$ have negative real parts, we have
that, for $\beta=0$, $x(t)=\bar{x}(t),\;y(t)=\bar{y}$ is stable as a
solution of (\ref{sync1sm}),(\ref{sync2sm}), hence by Lemma
\ref{pertn} $x(\beta,t)$,$y(\beta,t)$ is a stable solution of
(\ref{sync1sm}),(\ref{sync2sm}) for $|\beta|<\beta_0$. Thus
condition (C2) of Theorem \ref{stas} holds for this solution.

On if $\bar{x}(t)$ is unstable as a periodic solution of
(\ref{b02}), then Lemma \ref{pertn} implies that (C2) does not hold
for $|\beta|<\beta_0$, hence by Theorem \ref{stas} the synchronized
oscillation is unstable as a solution of (\ref{s1}),(\ref{s2}), so
that we have part (iv) of the Theorem.

To prove part (iii), we need to determine the circumstances under
which condition (C1) of Theorem \ref{stas} holds for the
synchronized oscillation when $|\beta|>0$ is small.

For the following computations it will be convenient to normalize
the period of the periodic solutions to $T_0$ by setting
\begin{equation}\label{dtau}\tau(\beta)=\frac{T(\beta)}{T_0},\end{equation}
$$u(\beta,t)=x(\beta,\tau(\beta) t),\;\;\;v(\beta,t)=y(\beta,\tau(\beta) t),$$ so that $u(\beta,t)$,
$v(\beta,t)$ are $T_0$-periodic with respect to $t$, and satisfy
\begin{equation}\label{ssync1}
u_t(\beta,t)=\tau(\beta)f(u(\beta,t),v(\beta,t)),
\end{equation}\begin{equation}\label{ssync2}
v_t(\beta,t)=\tau(\beta)[ g(v(\beta,t))+\beta
h(u(\beta,t),v(\beta,t))].
\end{equation}

To verify condition (C1) of Theorem \ref{stas} we need, defining
\begin{equation}\label{dc1}
a(\beta,t)=\tau(\beta)f_x(u(\beta,t),v(\beta,t)),
\end{equation}
to check whether all $d$ Floquet multipliers of the $T_0$-periodic linear equation
\begin{equation}\label{pse0s}
\dot{w}=a(\beta,t)w,
\end{equation}
have absolute value less than $1$.

Setting $\beta=0$ in (\ref{dc1}) we have
$a(0,t)=f_{x}(\bar{x}(t),\bar{y})$, so in this case (\ref{pse0s})
reduces to
\begin{equation}\label{rom}
\dot{w}=f_x(\bar{x}(t),\bar{y})w.
\end{equation}
(C1) does {\it{not}} hold for $\beta=0$, since (\ref{rom}) has the
Floquet multiplier $\mu=1$, corresponding to the $T_0$-periodic
solution $w(t)=\dot{\bar{x}}(t)$. However, by our assumption that
$\bar{x}(t)$ is a stable solution of (\ref{b02}), all {\it{other}}
Floquet multipliers of (\ref{rom}) are smaller than $1$ in absolute
value, and by continuity this remains true for (\ref{pse0s}) when
$|\beta|>0$ is sufficiently small. Our task, then, is to determine
in what way the Floquet multiplier  $\mu=1$ for $\beta=0$ is
perturbed when $|\beta|>0$ is small. Stability of the synchronized
oscillation corresponds to the perturbed Floquet multiplier being
inside the unit disk of the complex plane. We thus assume that the
Floquet multiplier $\mu=1$ of (\ref{pse0s}) is perturbed to
$\mu(\beta)$ for $|\beta|> 0$, where $\mu(\beta)$ is a real-valued
smooth function of $\beta$ with $\mu(0)=1$ - the justification for
which is the well-known lemma on perturbation of simple eigenvalues
(see e.g. \cite{kato}, Theorem 5.4). Thus for $|\beta|$ small there
is a solution $w(\beta,t)$ of
$$w_t(\beta,t)=
a(\beta,t)w(\beta,t),$$ satisfying
$$w(\beta,t+T_0)=\mu(\beta)w(\beta,t),\;\;\;\forall t\in\Real,$$
$$w(0,t)=\dot{\bar{x}}(t).$$
Defining
$$\eta(\beta)=\log(\mu(\beta)),$$
$$\tilde{w}(\beta,t)=e^{-\eta(\beta)t}w(\beta,t)$$
we have that
$$\eta(0)=0,$$
and $\tilde{w}(\beta,t)$ satisfies
\begin{equation}\label{weq}\tilde{w}_t(\beta,t)=
[a(\beta,t)-\eta(\beta)I]\tilde{w}(\beta,t),
\end{equation}
\begin{equation}\label{pew}
\tilde{w}(\beta,t+T_0)=\tilde{w}(\beta,t),\;\;\;\forall t\in\Real,
\end{equation}
\begin{equation}\label{rew}\tilde{w}(0,t)=\dot{\bar{x}}(t).\end{equation}
Stability for small $\beta>0$ will hold if $\eta'(0)<0$, which will
imply that $\eta(\beta)<0$ and hence $|\mu(\beta)|<1$. Similarly,
stability for small $\beta<0$ will hold if $\eta'(0)>0$. Part (iii)
of the Theorem will be proved by showing that
\begin{equation}\label{kr}
\eta'(0)=-\sigma,
\end{equation}
where $\sigma$ is given by (\ref{dsig}) and the rest of the proof is
devoted to this computation of $\eta'(0)$.

Differentiating (\ref{weq}) with respect to $\beta$ we have
\begin{align}\tilde{w}_{\beta t}(\beta,t)=[a_{\beta}(\beta,t)
-\eta'(\beta)I]\tilde{w}(\beta,t)\nonumber\\
+[a(\beta,t)-\eta(\beta)I]\tilde{w}_{\beta}(\beta,t)\nonumber
\end{align}
and putting $\beta=0$ and rearranging we get
\begin{align}\label{pet}\tilde{w}_{\beta t}(0,t)-f_{x}(\bar{x}(t),\bar{y})\tilde{w}_{\beta}(0,t)=
[a_{\beta}(0,t)-\eta'(0)I]\dot{\bar{x}}(t) .\end{align} Taking the
inner product of both sides of (\ref{pet}) with $q(t)$ (defined as
the $T_0$-periodic solution of (\ref{defq})) and integrating over
$[0,T_0]$, noting that, using integration by parts and (\ref{defq})
we have
\begin{align}\int_0^T \langle \tilde{w}_{\beta t}(0,s)-f_{x}(\bar{x}(s),\bar{y})\tilde{w}_{\beta}(0,s),q(s)\rangle ds\nonumber\\=
-\int_0^T \langle \tilde{w}_{\beta}(0,s),
\dot{q}(t)+[f_{x}(\bar{x}(s),\bar{y})]^*q(s) \rangle ds=0,
\end{align}
we get \begin{equation}\label{nn1}\int_0^{T_0}\langle
[a_{\beta}(0,s)-\eta'(0)I]\dot{\bar{x}}(s),q(s)\rangle
ds=0.\end{equation} Using (\ref{norm}) and (\ref{nn1}) we get
\begin{equation}\label{ett0}\eta'(0)=\int_0^{T_0}\langle a_{\beta}(0,s)\dot{\bar{x}}(s),q(s)\rangle ds
.\end{equation}
From (\ref{dc1}) we have
$$a(\beta,t)\dot{\bar{x}}(t)=\tau(\beta)f_x(u(\beta,t),v(\beta,t))\dot{\bar{x}}(t),$$
and differentiating this with respect to $\beta$ we get,
\begin{align}\label{abt}a_{\beta}(\beta,t)\dot{\bar{x}}(t)=\tau'(\beta)
f_x(u(\beta,t),v(\beta,t))\dot{\bar{x}}(t)\nonumber\\
+\tau(\beta) f_{xx}(u(\beta,t),v(\beta,t))[u_{\beta}(\beta,t),\dot{\bar{x}}(t)]\nonumber\\
+\tau(\beta)
f_{xy}(u(\beta,t),v(\beta,t))[\dot{\bar{x}}(t),v_{\beta}(\beta,t)].
\end{align}
Putting $\beta=0$ in (\ref{abt}) we have
\begin{align}\label{abt1}a_{\beta}(0,t)\dot{\bar{x}}(t)=\tau'(0)\ddot{\bar{x}}(t)
+ f_{xx}(\bar{x}(t),\bar{y})[u_{\beta}(0,t),\dot{\bar{x}}(t)]\nonumber\\
+
f_{xy}(\bar{x}(t),\bar{y})[\dot{\bar{x}}(t),v_{\beta}(0,t)].\end{align}

Differentiating (\ref{ssync1}) with respect to $\beta$ we have
\begin{eqnarray}\label{ubt}u_{\beta t}(\beta,t)&=&\tau'(\beta)\dot{\bar{x}}(t)
+\tau(\beta) f_x(u(\beta,t),v(\beta,t))
u_{\beta}(\beta,t)\nonumber\\
&+&\tau(\beta) f_y(u(\beta,t),v(\beta,t))v_{\beta}(\beta,t),
\end{eqnarray}
and setting $\beta=0$ in (\ref{ubt}) we get
\begin{align}\label{ut0}u_{\beta t}(0,t)=\tau'(0)\dot{\bar{x}}(t)+
 f_x(\bar{x}(t),\bar{y})u_{\beta}(0,t)
+  f_y(\bar{x}(t),\bar{y})v_{\beta}(0,t).\end{align}

Differentiating (\ref{ut0}) with respect to $t$ we get
\begin{eqnarray}\label{ubtt}u_{\beta tt}(0,t)&=&\tau'(0)\ddot{\bar{x}}(t)
+f_{xx}(\bar{x}(t),\bar{y})[\dot{\bar{x}}(t),u_{\beta}(0,t)]+
 f_x(\bar{x}(t),\bar{y})u_{\beta t}(0,t)\nonumber\\
&+&  f_{xy}(\bar{x}(t),\bar{y})[\dot{\bar{x}}(t),v_{\beta}(0,t)]+
f_{y}(\bar{x}(t),\bar{y})v_{\beta t}(0,t).
\end{eqnarray}
Combining (\ref{abt1}) and (\ref{ubtt}) we get
\begin{align}a_{\beta}(0,t)\dot{\bar{x}}(t)=u_{\beta tt}(0,t)- f_x(\bar{x}(t),\bar{y})
u_{\beta t}(0,t)- f_{y}(\bar{x}(t),\bar{y})v_{\beta
t}(0,t),\nonumber
\end{align}
which we can also write as
\begin{align}\label{tar}a_{\beta}(0,t)\dot{\bar{x}}(t)+ f_{y}(\bar{x}(t),\bar{y})v_{\beta t}(0,t)
=u_{\beta tt}(0,t)-
 f_{x}(\bar{x}(t),\bar{y})u_{\beta t}(0,t).\end{align}

Taking the inner product of (\ref{tar}) with $q(t)$ and integrating
over $[0,T_0]$, and noting that, using (\ref{defq}), the right-hand
side vanishes, we get
\begin{align}\int_0^{T_0}\langle a_{\beta}(0,s)\dot{\bar}{x}(s),q(s)\rangle ds
+ \int_0^{T_0}\langle f_{y}(\bar{x}(s),\bar{y})v_{\beta
t}(0,s),q(s)\rangle ds=0,\nonumber\end{align} which, together with
(\ref{ett0}), gives
\begin{align}\label{e1}\eta'(0)=
- \int_0^{T_0}\langle f_{y}(\bar{x}(s),\bar{y})v_{\beta
t}(0,s),q(s)\rangle ds.
\end{align}
We now compute $v_{\beta}(0,t)$. Differentiating (\ref{ssync2}) with
respect to $\beta$ and setting $\beta=0$, we have
\begin{equation}\label{ydif}v_{\beta t}(0,t)- Av_{\beta}(0,t)=
h(\bar{x}(t),\bar{y}).
\end{equation}
Solving (\ref{ydif}) for $v_{\beta}(0,t)$, we get
\begin{equation}\label{solvv}v_{\beta}(0,t) =e^{tA}v_{\beta}(0,0)+\int_{0}^{t}
e^{rA}h(\bar{x}(t-r),\bar{y})dr,\end{equation} and using the fact
that $v_{\beta}(0,t)$ and $\bar{x}(t)$ are $T_0$-periodic we have
\begin{equation}\label{v00}v_{\beta}(0,0) =[I-e^{T_0A}]^{-1}\int_{0}^{T_0}
e^{rA}h(\bar{x}(-r),\bar{y})dr.\end{equation}
We have
\begin{eqnarray*}&&\int_{0}^{\infty} e^{rA}h(\bar{x}(-r),\bar{y})dr=\sum_{k=0}^{\infty}
\int_{0}^{T_0} e^{(r+kT_0)A}h(\bar{x}(kT_0-r),\bar{y})dr
\\&=&\Big[\sum_{k=0}^{\infty}e^{kT_0 A}\Big]
\int_{0}^{T_0}
e^{rA}h(\bar{x}(-r),\bar{y})dr=[I-e^{T_0A}]^{-1}\int_{0}^{T_0}
e^{rA}h(\bar{x}(-r),\bar{y})dr,\end{eqnarray*}
 where we have used
the periodicity of $\bar{x}(t)$, and the fact that the geometric
series converges because the eigenvalues of $A$ have negative real
parts. Therefore (\ref{v00}) can be rewritten as
\begin{equation}\label{v01}v_{\beta}(0,0) =\int_{0}^{\infty} e^{rA}h(\bar{x}(-r),\bar{y})dr.\end{equation}
From (\ref{solvv}) and (\ref{v01}) we have
\begin{eqnarray}\label{vb}v_{\beta}(0,t)
&=&e^{tA}\int_{0}^{\infty} e^{rA}h(\bar{x}(-r),\bar{y})dr
+\int_{0}^{t} e^{rA}h(\bar{x}(t-r),\bar{y})dr\nonumber\\
&=&\int_{t}^{\infty} e^{rA}h(\bar{x}(t-r),\bar{y})dr +\int_{0}^{t}
e^{rA}h(\bar{x}(t-r),\bar{y})dr\nonumber\\ &=&\int_{0}^{\infty}
e^{rA}h(\bar{x}(t-r),\bar{y})dr,
\end{eqnarray}
and, using (\ref{ydif}) and (\ref{vb}),
\begin{eqnarray}\label{ydif1}&&v_{\beta t}(0,t)=h(\bar{x}(t),\bar{y})+Av_{\beta}(0,t)\\
&=&h(\bar{x}(t),\bar{y})
+A\int_{0}^{\infty} e^{rA}h(\bar{x}(t-r),\bar{y})dr\nonumber\\
&=&A\int_{0}^{\infty} e^{rA}[h(\bar{x}(t-r),\bar{y})-h(\bar{x}(t),\bar{y})]dr\nonumber\\
&=&\int_{0}^{\infty} \frac{d}{dr}[e^{rA}][h(\bar{x}(t-r),\bar{y})-h(\bar{x}(t),\bar{y})]dr\nonumber\\
&=&e^{rA}[h(\bar{x}(t-r),\bar{y})-h(\bar{x}(t),\bar{y})]\Big|_{r=0}^{r=\infty}-\int_{0}^{\infty}
e^{rA}\frac{d}{dr}h(\bar{x}(t-r),\bar{y})dr\nonumber\\
&=&\int_{0}^{\infty}
e^{rA}h_x(\bar{x}(t-r),\bar{y})\dot{\bar{x}}(t-r) dr.\nonumber
\end{eqnarray}
Substituting the expression (\ref{ydif1}) for $v_{\beta t}(0,s)$
into (\ref{e1}), we get (\ref{kr}), as we wanted. We now prove part
(ii). We rewrite (\ref{ut0}) as
\begin{align}\label{ut1}u_{\beta t}(0,t)-f_{x}(\bar{x}(t),\bar{y})u_{\beta}(0,t)=\tau'(0)\dot{\bar{x}}(t)
+ f_y(\bar{x}(t),\bar{y})(v_{\beta}(0,t)).\end{align} Taking the
inner product of (\ref{ut1}) with $q(t)$ and integrating over
$[0,T_0]$ using the fact that, due to (\ref{defq}), the left-hand
side vanishes and (\ref{norm}), we get
\begin{align}\label{last}
\tau'(0) =-  \int_0^{T_0} \langle
f_y(\bar{x}(s),\bar{y})v_{\beta}(0,s),q(s)\rangle ds.
\end{align}
Substituting the expression (\ref{vb}) for $v_{\beta}(0,s)$ into
(\ref{last}), we get
$$\tau'(0)
 =-  \int_{0}^{\infty}\int_0^{T_0} \langle
f_y(\bar{x}(s),\bar{y})e^{rA}h(\bar{x}(s-r),\bar{y}), q(t)\rangle ds
dr$$ which, in view of (\ref{dtau}) implies (\ref{pers}).
\end{proofw}

\section{Application to a model for the pulsatile secretion of GnRH}
\label{gonadotropin}

In this section we apply Theorem \ref{stas} to study a mathematical model, presented
by Khadra and Li \cite{khadra}, whose aim is to explain the synchronization of
the periodic (with period approximately $1$ hour) secretion of GnRH (gonadotropin-releasing hormone) by
GnRH neurons in the hypothalamus. The explanation proposed for this synchronization phenomenon, based
on a range of experimental results, is
that the GnRH neurons have receptors for GnRH, so that the concentration of GnRH in the environment
influences their dynamics by binding to these receptors,
thus inducing an indirect coupling leading to synchronization.

We very briefly describe the mechanisms involved in the model of
\cite{khadra}, and refer to that paper for details. The binding of
GnRH to the receptors on a GnRH neuron activates three types of
G-protein $G_s,G_q,G_i$ in the cell, and the concentrations of their
activated subunits $\alpha_s,\alpha_q,\alpha_i$ in the cell are
denoted by $S,Q,I$, respectively. $\alpha_s$ activates the
production of $cAMP$, whose concentration is denoted by $A$.
$\alpha_q$ induces the release of ${\mbox{Ca}}^{2+}$, whose
concentration is denoted by $C$, from intracellular stores. $C$ and
$A$ act in synergy to induce the secretion of GnRH from the neuron.
$G$ denotes the concentration of GnRH in the external environment.
These causal relations are modeled by the following differential
equations:
\begin{equation}\label{gon11}
\dot{S}=\nu_S H_S(G)-k_S S,
\end{equation}
\begin{equation}\label{gon21}
\dot{Q}=\nu_Q H_Q(G)-k_Q Q,
\end{equation}
\begin{equation}\label{gon31}
\dot{I}=\nu_I H_S(G)-k_I I,
\end{equation}
\begin{equation}\label{gon41}
\dot{C}=J_{IN}+[l+\nu_C F_C(C,Q)](C_{ER}-C)-k_C C,
\end{equation}
\begin{equation}\label{gon51}
\dot{A}=b_A+\nu_A F_A(S,I)-k_A A,
\end{equation}
\begin{equation}\label{gon61}
\dot{G}=b_G+\nu_G F_G(C,A)-k_G G,
\end{equation}
where all the parameters, and the nonlinearities $H_S,H_Q,H_I,F_C,F_A,F_G$, are positive.
In \cite{khadra}  these nonlinearities are taken as:
\begin{align}\label{sp}
H_{S}(G)=\frac{G^4}{K_S^4+G^4},\;\;H_{Q}(G)=\frac{G^2}{K_Q^2+G^2},\;\;H_{I}(G)=\frac{G^2}{K_I^2+G^2}.\nonumber\\
F_C(C,Q)=Q,\;\;F_A(S,I)=\frac{h_1 S}{I+h_1},\;\;F_G(C,A)=(AC)^3.
\end{align}
For our results we do not need to assume these specific forms.

When dealing with $n$ GnRH neurons coupled through the GnRH in their environment, the
model becomes \cite{khadra} ($1\leq k\leq n$)
\begin{equation}\label{gon1}
\dot{S}_k=\nu_S H_S(G)-k_S S_k,
\end{equation}
\begin{equation}\label{gon2}
\dot{Q}_k=\nu_Q H_Q(G)-k_Q Q_k,
\end{equation}
\begin{equation}\label{gon3}
\dot{I}_k=\nu_I H_S(G)-k_I I_k,
\end{equation}
\begin{equation}\label{gon4}
\dot{C}_k=J_{IN}+[l+\nu_C F_C(C_kQ_k)](C_{ER}-C_k)-k_C C_k,
\end{equation}
\begin{equation}\label{gon5}
\dot{A}_k=b_A+\nu_A F_A(S_k,I_k)-k_A A_k,
\end{equation}
\begin{equation}\label{gon6}
\dot{G}=b_G+\frac{\nu_G}{n} \sum_{j=1}^n F_G(C_j,A_j)-k_G G,
\end{equation}
where $S_k,Q_k,I_k,I_k,C_k,A_k$ are intracellular concentrations of the various species in neuron $k$, and
$G$ is the concentration of GnRH in the intercellular medium.

We are going to prove results which say that if a single neuron,
placed in the environment, performs periodic oscillations (in other
words if the system (\ref{gon11})-(\ref{gon61}) has a stable
periodic solution), then a population of such neurons will
synchronize (in the sense that the synchronized oscillation is
stable).

The following simple lemma will be used.
\begin{lemma}\label{sim}
Assume $a(t)$ satisfies
$$a(t)\geq a_*>0,\;\;\forall t>0,$$
and $f(t)$ satisfies
\begin{equation}\label{q1}\dot{f}(t)+a(t)f(t)=y(t)\end{equation}
where
\begin{equation}\label{q2}|y(t)|\leq m e^{-kt}\;\;\;\forall t>0\end{equation}
with $k>0$.
Then $f(t)$ converges exponentially to $0$ as $t\rightarrow \infty$: there exist $m',k'>0$ so that
$$|f(t)|\leq m' e^{-k't}\;\;\forall t>0.$$
\end{lemma}
\begin{proof}
$f$ can be written explicitly as
\begin{eqnarray*}
f(t)&=&\exp \Big(-\int_0^t a(s)ds \Big)f(0)+ \int_0^t
\exp \Big(-\int_s^t a(r)dr \Big)y(s) ds.\nonumber
\end{eqnarray*}
Therefore, using (\ref{q1}),(\ref{q2}) we have
\begin{eqnarray*}
|f(t)|&\leq &\exp (-t a_* )|f(0)|+ me^{-a_*t}\int_0^t
e^{(a_*-k)s} ds.\nonumber\\
&=&\exp (-t a_* )|f(0)|+m\frac{e^{-kt}-e^{-a_*t}}{a_*-k},
\end{eqnarray*}
which gives the exponential decay.
\end{proof}

We assume that the system
(\ref{gon11})-(\ref{gon61})
has a stable periodic solution $S(t)$, $Q(t)$, $I(t)$, $C(t)$, $A(t)$, $G(t)$ (that is, we assume that
a single neuron performs oscillations).
Condition (C2) of Theorem \ref{stas} holds by this assumption.
To obtain the stability of this solution as a synchronized oscillation of (\ref{gon1})-(\ref{gon6}),
we need, according to Theorem \ref{stas}, to verify (C1), that is to show that the Floquet multipliers
of the periodic equation system
\begin{equation}\label{lgon1}
\dot{\tilde{S}}=-k_S \tilde{S}
\end{equation}
\begin{equation}\label{lgon2}
\dot{\tilde{Q}}=-k_Q \tilde{Q}
\end{equation}
\begin{equation}\label{lgon3}
\dot{\tilde{I}}=-k_I \tilde{I}
\end{equation}
\begin{eqnarray}\label{lgon4}
\dot{\tilde{C}}&=&-\Big[l+k_C+\nu_C F_C(C(t),Q(t))+\frac{\partial
F_C}{\partial C}(C(t),Q(t))(C_{ER}-C(t))\Big]
\tilde{C}\nonumber\\
&+&\frac{\partial F_C}{\partial Q}(C(t),Q(t))(C_{ER}-C(t))\tilde{Q},
\end{eqnarray}
\begin{equation}\label{lgon5}
\dot{\tilde{A}}=\nu_A \frac{\partial F_A}{\partial
S}(S(t),I(t))\tilde{S} +\nu_A \frac{\partial F_A}{\partial
I}(S(t),I(t))\tilde{I}-k_A \tilde{A}.
\end{equation}
have absolute values less than $1$, or in other words that
any solution of this system decays to $0$ at an exponential rate as $t\rightarrow \infty$.
Let $\tilde{S}(t)$, $\tilde{Q}(t)$, $\tilde{I}(t)$, $\tilde{C}(t)$, $\tilde{A}(t)$,
be a solution of (\ref{lgon1})-(\ref{lgon5}).
From (\ref{lgon1})-(\ref{lgon3}) we get that
\begin{equation}\label{eb1}
\tilde{S}(t)= \tilde{S}(0)e^{-k_S t}
,\;\;\;
\tilde{Q}(t)= \tilde{Q}(0)e^{-k_Q t},
\;\;\;
\tilde{I}(t)=\tilde{I}(0)e^{-k_I t},
\end{equation}
so that these components certainly decay exponentially.

Substituting (\ref{eb1}) into (\ref{lgon5}) we have
\begin{equation}\label{lgon53}\dot{\tilde{A}}+k_A \tilde{A}=\nu_A \frac{\partial F_A}{\partial S}(S(t),I(t))\tilde{S}(0)e^{-k_S t}
+\nu_A \frac{\partial F_A}{\partial I}(S(t),I(t))\tilde{I}(0)e^{-k_I t}.\end{equation}
The right-hand side of (\ref{lgon53}) decays exponentially, and $k_A>0$,
hence, by Lemma \ref{sim}, $A(t)$ decays exponentially.

Substituting (\ref{eb1}) into (\ref{lgon4})
we get
\begin{align}\label{rgon4}
\dot{\tilde{C}}+K(t)\tilde{C}= \frac{\partial F_C}{\partial
Q}(C(t),Q(t))(C_{ER}-C(t))\tilde{Q}(0)e^{-k_Q t},
\end{align}
where
\begin{equation}\label{defk}K(t)=l+k_C+\nu_C F_C(C(t),Q(t))+\frac{\partial F_C}{\partial C}(C(t),Q(t))(C_{ER}-C(t)).
\end{equation}
The right-hand side of (\ref{defk}) decays exponentially, so
to use Lemma \ref{sim} in order to show that $\tilde{C}(t)$ decays exponentially, we must show that
\begin{equation}\label{kp}\min_{t\in\Real}K(t)>0.
\end{equation}

If we assume that $F_C$ does not depend on $C$, so that
\begin{equation}\label{zz}
\frac{\partial F_C}{\partial C}(C,Q)=0,
\end{equation}
which holds in the case of (\ref{sp}), then
$$K(t)=l+k_C+\nu_C F_C(C(t),Q(t))\geq l+k_C>0,$$
hence (\ref{kp}) holds. We thus have

\begin{proposition}\label{pr1}
Assume that (\ref{zz}) holds and that the system
(\ref{gon11})-(\ref{gon61})
has a stable periodic solution $S(t)$, $Q(t)$, $I(t)$, $C(t)$, $A(t)$, $G(t)$.
Then, for any $n\geq 1$,  the synchronized
 oscillation $S_k(t)=S(t)$, $Q_k(t)=Q(t)$, $I_k(t)=I(t)$, $A_k(t)=A(t)$ ($1\leq k\leq n$), $G(t)$
is a stable solution of (\ref{gon1})-(\ref{gon6}).
\end{proposition}

Proposition \ref{pr1} covers the case in which the nonlinearities
are given by (\ref{sp}), but we now also derive a sufficient
condition for synchronization without assuming (\ref{zz}). As
explained in \cite{khadra} there is evidence for positive feedback
of $Ca^{2+}$ concentration on $Ca^{2+}$ release, so that we assume
\begin{equation}\label{bbb1}
\frac{\partial F_C}{\partial C}(C,Q)\geq 0.
\end{equation}

We note that (\ref{gon41}) implies that
\begin{eqnarray*}
\dot{C}(t)=0\;\; &\Rightarrow&\;\;C(t)= \frac{J_{IN}+[l+\nu_C F_C(C(t),Q(t))]C_{ER}}{k_C +l+\nu_C F_C(C(t),Q(t))}\\
\end{eqnarray*}
and since
\begin{eqnarray*}
0<\frac{J_{IN}+[l+\nu_C F_C(C(t),Q(t))]C_{ER}}{k_C +l+\nu_C F_C(C(t),Q(t))}\leq
\max \Big( \frac{J_{IN}+lC_{ER}}{k_C +l},
C_{ER} \Big),
\end{eqnarray*}
we have that if $\dot{C}(t)=0$ then
\begin{equation}\label{c0}
0< C(t)\leq
\max \Big( \frac{J_{IN}+lC_{ER}}{k_C +l},
C_{ER} \Big),
\end{equation}
so that in particular (\ref{c0}) holds at the minimum and maximum points of $C(t)$ (recall
that $C(t)$ is periodic), hence (\ref{c0}) holds for all $t$.
If we assume that
$$C_{ER}>\frac{J_{IN}+lC_{ER}}{k_C +l}
,$$
then we get $C(t)<C_{ER}$ for all $t$, hence from (\ref{defk}) and (\ref{bbb1}) we get that (\ref{kp}) holds.
We thus obtain
\begin{proposition}\label{sec}
Assume that (\ref{bbb1}) and
\begin{equation}\label{hg}C_{ER}k_C >J_{IN}\end{equation}
 hold, and that the system
(\ref{gon11})-(\ref{gon61})
has a stable periodic solution $S(t)$, $Q(t)$, $I(t)$, $C(t)$, $A(t)$, $G(t)$.
Then, for any $n\geq 1$,  the synchronized
 oscillation $S_k(t)=S(t)$, $Q_k(t)=Q(t)$, $I_k(t)=I(t)$, $A_k(t)=A(t)$ ($1\leq k\leq n$), $G(t)$
is a stable solution of (\ref{gon1})-(\ref{gon6}).
\end{proposition}

The biologically reasonable values of the parameters given in \cite{khadra},
$C_{ER}=2.5\mu M$, $k_C=5100 \min^{-1}$, $J_{IN}=0.2 \frac{\mu M}{\min}$, are well within the
range satisfying (\ref{hg}), so that proposition \ref{sec} applies and assures that synchronization
will occur, even if the nonlinearity $F_C$ depends on both $Q$ and $C$.

We remark that Khadra and Li also give a simplified version of their model (\cite{khadra}, eqs. 8-9).
Proving the stability of synchronized oscillations of this simplified model (assuming that
a single cell has a stable oscillation), by verifying condition (C1) of \ref{stas}, is even easier than in the case of the full
model, and no restriction on the parameters or nonlinearities is needed in this case.

%

\section{Synchronization of indirectly coupled $\lambda-\omega$ oscillators}
\label{hopf}

$\lambda-\omega$ oscillators are given by the equations
\begin{align}\label{lw}
\dot{u}=\lambda(\sqrt{u^2+v^2})u-\omega(\sqrt{u^2+v^2})v\nonumber\\
\dot{v}=\omega(\sqrt{u^2+v^2})u+\lambda(\sqrt{u^2+v^2})v,
\end{align}
where $\omega,\lambda: [0,\infty)\rightarrow \Real$ are given functions.
Introducing polar coordinates $r,\theta$ in the $u,v$-plane, we can write (\ref{lw}) as
$$\dot{r}=\lambda(r)r,\;\;\;\dot{\theta}=\omega(r).
$$
It is then seen immediately that if $r_0>0$ is such that
$\lambda(r_0)=0$ then
$$\bar{u}(t)=r_0\cos(\omega_0t),\;\;\bar{v}(t)=r_0\sin(\omega_0t),$$
where $\omega_0=\omega(r_0)$. Moreover, if $\lambda'(r_0)<0$ then this
periodic solution is stable, and if $\lambda'(r_0)>0$ it is unstable.

In the particular case
$$\lambda(r)=1-r^2,\;\;\omega(r)=1 +\gamma(1- r^2)$$
we get the Ginzburg-Landau oscillator, which in terms of $A=u+iv$ can be written as
\begin{equation}\label{gl}\dot{A}=(1+i(\gamma+1)) A-(1+i \gamma)|A|^2 A.\end{equation}
When $\gamma=0$ this is known as the `radial isochron clock'
\cite{hop}, or `Poincar\'e oscillator' \cite{glass}.

We shall assume that
$$\lambda(1)=0,\;\;\;\lambda'(1)<0,$$
$$\omega(1)=1.$$
so that (\ref{lw}) has the  stable periodic solution
$$\bar{u}(t)=\cos(t),\;\;\bar{v}(t)=\sin(t).$$
The existence of an explicitly-known periodic solution facilitates
analytical study of weakly-coupled $\lambda-\omega$ oscillators without resort to numerical
calculations. $\lambda-\omega$ oscillators have been used to address a variety of
questions related to coupled and forced oscillators - see \cite{winfree}, page 163 for references.

We will use Theorem \ref{existence} to study a system of $n$
indirectly coupled $\lambda-\omega$ oscillators ($1\leq k \leq n$)
\begin{equation}\label{cv1}\dot{u}_k=\lambda\Big(\sqrt{u_k^2+v_k^2}\Big)u_k
-\omega\Big(\sqrt{u_k^2+v_k^2}\Big)v_k+F(u_k,v_k)y,\end{equation}
\begin{equation}\label{cv2}\dot{v}_k=\omega\Big(\sqrt{u_k^2+v_k^2}\Big)u_k+\lambda\Big(\sqrt{u_k^2+v_k^2}\Big)v_k,
\end{equation}
\begin{equation}\label{cv3}\dot{y}=-\alpha y+\frac{\beta}{n}\sum_{j=1}^n (a u_j +b v_j),\end{equation}
where $\alpha>0$, in the weakly coupled case $|\beta|\rightarrow 0$.
We will obtain conditions on the function $F$ and on the parameters
$\alpha,a,b$ which ensure that the synchronized oscillation, which
exists for $|\beta|$ small, is (un)stable.

The system (\ref{cv1})-(\ref{cv3}) is of the form (\ref{s1}),(\ref{s2}), with $d=2$, $p=1$,
$$x=\left(
      \begin{array}{c}
        u \\
        v \\
      \end{array}
    \right),$$

$$f(x
,y)=\left(
      \begin{array}{c}
       \lambda(\sqrt{u^2+v^2})u-\omega(\sqrt{u^2+v^2})v+F(u,v)y \\
       \omega(\sqrt{u^2+v^2})u+\lambda(\sqrt{u^2+v^2})v \\
      \end{array}
    \right),
$$
$$g(y)=-\alpha y$$
$$h(x
,y)=au+bv,$$
When $\beta=0$, the uncoupled system satisfies Assumption \ref{assume1}: the equation
(\ref{b01}) has the stable stationary solution $\bar{y}=0$, and the equation (\ref{b02})
is the $\lambda-\omega$ oscillator, with the solution
$$\bar{x}(t)=\left(
               \begin{array}{c}
                \cos(t) \\
                \sin(t) \\
               \end{array}
             \right).
$$
We have \begin{eqnarray*}&&
f_x(\bar{x}(t),0)\\&=&\begin{footnotesize}\left(
                      \begin{array}{cc}
                        \cos(t)[\lambda'(1)\cos(t)-\omega'(1)\sin(t)]  & \sin(t)[\lambda'(1)\cos(t)-\omega'(1)\sin(t)]-1\\
                        \cos(t)[\omega'(1)\cos(t)+\lambda'(1)\sin(t)]+1 & \sin(t)[\omega'(1)\cos(t)+\lambda'(1)\sin(t)] \\
                      \end{array}
                    \right).\end{footnotesize}
\end{eqnarray*}
One can check by inspection that
$$q(t)=\left(
         \begin{array}{c}
           q_1(t) \\
           q_2(t) \\
         \end{array}
       \right)=\frac{1}{2\pi}\left(
         \begin{array}{c}
           \lambda'(1)\sin(t)+\omega'(1)\cos(t) \\
           -\lambda'(1)\cos(t)+\omega'(1)\sin(t) \\
         \end{array}
       \right),
$$
is the $2\pi$-periodic solution of the equation (\ref{defq}),
where the coefficient $\frac{1}{2\pi}$ is taken to achieve the normalization (\ref{norm}).

We compute also
$$f_y(x,y)=\left(
             \begin{array}{c}
               F(u,v) \\
               0 \\
             \end{array}
           \right)
$$
$$h_x(x,y)=\left(
              \begin{array}{cc}
                a & b \\
              \end{array}
            \right)
$$
\begin{align}f_y(\bar{x}(s),0) e^{rA}h_x(\bar{x}(s-r),
0)
=e^{-\alpha y}\left(
     \begin{array}{cc}
       aF(\bar{u}(s),\bar{v}(s)) & bF(\bar{u}(s),\bar{v}(s)) \\
       0 & 0 \\
     \end{array}
   \right),\nonumber
\end{align}
\begin{eqnarray*}&&f_y(\bar{x}(s),0) e^{rA}h_x(\bar{x}(s-r),
0)\dot{\bar{x}}(s-r)\nonumber\\
&=& e^{-\alpha r}\left(
                   \begin{array}{c}
                   aF(\bar{u}(s),\bar{v}(s))\dot{\bar{u}}(s-r)+bF(\bar{u}(s),\bar{v}(s))\dot{\bar{v}}(s-r)\\
                   0\\
                   \end{array}
                 \right)
\nonumber
\end{eqnarray*}
\begin{eqnarray*}&&\langle f_y(\bar{x}(s),0) e^{rA}h_x(\bar{x}(s-r),
0)\dot{\bar{x}}(s-r),q(s)\rangle\nonumber\\
&=&e^{-\alpha r}[aF(\bar{u}(s),\bar{v}(s))\dot{\bar{u}}(s-r)+bF(\bar{u}(s),\bar{v}(s))\dot{\bar{v}}(s-r)]q_1(s)\nonumber\\
&=&\frac{e^{-\alpha r}}{2\pi} [-aF(\cos(s),\sin(s))\sin(s-r)
+bF(\cos(s),\sin(s))\cos(s-r)]\\&\times& [\lambda'(1)\sin(s)+\omega'(1)\cos(s)],
\nonumber\end{eqnarray*}
hence
\begin{eqnarray*}&&
\sigma=\\&&\frac{b}{2\pi}\int_0^{2\pi}F(\cos(s),\sin(s))[\lambda'(1)\sin(s)+
\omega'(1)\cos(s)]\int_0^{\infty}\cos(s-r)e^{-\alpha r}dr ds\\
&-&\frac{a}{2\pi}\int_0^{2\pi}F(\cos(s),\sin(s))[\lambda'(1)\sin(s)+\omega'(1)\cos(s)]\int_0^{\infty}\sin(s-r)e^{-\alpha r}dr ds
\\&=&\frac{1}{\alpha^2+1}\Big[(b-\alpha a) [\omega'(1)C_2+\lambda'(1)C_1]
+(a+\alpha b) [\omega'(1)C_3+\lambda'(1)C_2]
\Big]
\end{eqnarray*}
where
\begin{align}\label{dc}C_1=\frac{1}{2\pi}\int_0^{2\pi}F(\cos(s),\sin(s))\sin^2(s)ds,\nonumber\\
C_2=\frac{1}{2\pi}\int_0^{2\pi}F(\cos(s),\sin(s))\cos(s)\sin(s)ds,\nonumber\\
C_3=\frac{1}{2\pi}\int_0^{2\pi}F(\cos(s),\sin(s))\cos^2(s)ds.
\end{align}

From Theorem \ref{existence} we thus obtain
\begin{proposition}
Assuming $\lambda'(1)\neq0$ and $\alpha\neq 0$, there exists
$\beta_0>0$ such that for $|\beta|< \beta_0$ (\ref{cv1})-(\ref{cv3})
has a synchronized oscillation, and we have

\noindent (i) If $\lambda'(1)<0$ and
\begin{equation}\label{scd}(b-\alpha a) [\omega'(1)C_2+\lambda'(1)C_1]
+(a+\alpha b) [\omega'(1)C_3+\lambda'(1)C_2]>0,\end{equation}
where $C_1,C_2,C_3$ are defined by (\ref{dc}),
then the synchronized oscillation is stable for $\beta>0$ and unstable for $\beta<0$.

\noindent (ii) If $\lambda'(1)<0$ and the reverse inequality to
(\ref{scd}) holds, then the
 synchronized oscillation is unstable for $\beta>0$ and stable for $\beta<0$.

\noindent (iii) If $\lambda'(1)>0$  then the synchronized
oscillation is unstable for all $|\beta|<\beta_0$.
\end{proposition}

In the case of coupled Ginzburg-Landau oscillators (\ref{gl}), where
we have $\lambda'(1)=-2$, $\omega'(1)=-2\gamma$, we get

\begin{corollary}
There exists $\beta_0>0$ such that
 for $|\beta|< \beta_0$
the system
of coupled Ginzburg-Landau oscillators ($1\leq k\leq n$):
\begin{equation*}\dot{u}_k=[1-(u_k^2+v_k^2)]u_k
-[1+\gamma-\gamma(u_k^2+v_k^2)]v_k+F(u_k,v_k)y,\end{equation*}
\begin{equation*}\dot{v}_k=[1+\gamma-\gamma(u_k^2+v_k^2)]u_k+[1-(u_k^2+v_k^2)]
v_k,\end{equation*}
\begin{equation*}\dot{y}=-\alpha y+\frac{\beta}{n}\sum_{j=1}^n (a u_j +b v_j),\end{equation*}
has a synchronized oscillation, which is stable if
\begin{equation}\label{fr}\beta[(b-\alpha a) (\gamma C_2+C_1)
+(a+\alpha b) (\gamma C_3+C_2)]<0,\end{equation} and unstable if the
reverse inequality holds, where $C_1,C_2,C_3$ are defined by
(\ref{dc}).
\end{corollary}

\section{Discussion}
\label{discussion}

We summarize here the basic insights provided by the analytical
results given by Theorems \ref{stas} and \ref{existence}, and make
some remarks about the possibilities for applying these results to
the study of specific systems.

Theorem \ref{stas} shows that the study of synchronization, that is
the determination of (in)stability of a synchronized solution of
(\ref{s1}),(\ref{s2}), which is a system of size $nd+p$, reduces to
the study of the study of two linear systems of dimensions $d$
(equation \ref{pse0}) and $d+p$ (equation \ref{eq2}), associated
with a periodic oscillation of a single oscillator. As we have
demonstrated in section \ref{gonadotropin}, for certain systems this
result can be used to prove synchronization without resort to
numerical computations, but in general this will not be the case.
However, as we have noted, Theorem \ref{stas} has the following
implication which is significant in general: if the system
(\ref{s1}),(\ref{s2}) with $n=2$ has a stable synchronized
oscillation, then the same is true for any $n$. If the system with
$n=2$ is studied by numerical simulation and observed to display
synchronization, then we are assured that the synchronized
oscillation will be stable for the system with arbitrarily large
$n$. The caveat must be made here that since the notion of stability
is a local one, it is possible that for $n=2$ the synchronized
oscillation will be globally stable, but for some larger $n$ the
synchronized oscillation will only be locally stable. Finding
criteria for global stability of the synchronized oscillation of
system (\ref{s1}),(\ref{s2}) is an interesting question for further
research.

Theorem \ref{existence} provides an understanding of synchronization
in the case of weak coupling ($|\beta|$ small). It is shown that
synchronization can occur for arbitrarily weak coupling, provided it
is of the `right' sign, as determined by the integral $\sigma$. It
is to be noted that in many cases only a positive value for $\beta$
makes physical sense, as in the case of coupled cells where $\beta$
is given by (\ref{dbeta}), in which case the condition for
synchronization is $\sigma<0$. This criterion can be used for a
systematic studies of synchronization in the weak-coupling regime,
in dependence on various parameters. An example of such a study was
given in section \ref{hopf} for $\lambda-\omega$ oscillators. In
this case calculations are particularly simple, because the periodic
solution in the uncoupled case is available in closed form. More
generally such a study can be performed with the aid of numerical
computation. Suppose that the coupling function $h$ in (\ref{s2})
depends on some parameters $\alpha=(\alpha_1,\ldots,\alpha_m)$, and
we want to determine the subset in the space of parameters $\alpha$
that will lead to synchronization for small $\beta>0$. We first
compute the periodic solution $\bar{x}(t)$ of (\ref{b02}), for
example by direct numerical simulation. We then substitute
$\bar{x}(t)$ into the formula (\ref{dsig}) for $\sigma$. The
dependence of $\sigma$ on $\alpha$ follows from $\alpha$-dependence
of $h$, and the function $\sigma(\alpha)$ can be computed by a
numerical integration. The surface $\sigma(\alpha)=0$ will separate
the parameter space into synchronizing and non-synchronizing regions
in the small parameter case.

%
%


\begin{thebibliography}{9}



\bibitem{camacho} E. Camacho, R. Rand \& H. Howland, {\it{Dynamics of two van der Pol oscillators coupled
via a bath}}, Int. J. Solids \& Structures {\bf{41}} (2004), 2133-2143.

\bibitem{foster} R. Foster, B. Kreitzman, `Rhythms of Life',
Profile Books (London), (2004).

\bibitem{furusawa} C. Furusawa \& K. Kaneko, {\it{Emergence of rules in cell society: differentiation, hierarchy and
stability}}, Bull. Math. Biol. {\bf{60}} (1998), 659-687.

\bibitem{garcia} J. Garcia-Ojalvo, M.B. Elowitz \& S.H. Strogatz,
{\it{Modeling a synthetic multicellular clock: Repressilators
coupled by quorum sensing}}, PNAS {\bf{27}} (2004), 10955-10960.

\bibitem{geier} F. Geier, S. Becker-Weimann, A. Kramer \& H. Herzel,
{\it{Entrainment in a model of the mamalian circadian oscillator}},
J. Biol. Rhythms {\bf{20}} (2005), 83-93.

\bibitem{glass} L. Glass \& M.C. Mackey, `From Clocks to Chaos: The Rhythms of Life',
Princeton U. Press (Princeton), 1988.

\bibitem{golbeter} A. Goldbeter, `Biochemical Oscillations and Cellular Rhythms', Cambridge U. Press (Cambridge), 1996.

\bibitem{gonze} D. Gonze, S. Bernard, C. Waltermann, A. Kramer \& H. Herzel,
{\it{Spontaneous synchronization of coupled circadian oscillators}},
Biophys. J. {\bf{89}} (2005), 120-129.

\bibitem{henson} M.A. Henson, {\it{Modeling the synchronization of yeast respiratory oscillations}},
J. Theo. Biol. {\bf{231}} (2004), 443-458.

\bibitem{hess} B. Hess, {\it{Periodic patterns in biochemical reactions}}, Quart. Rev. Biophys. {\bf{30}} (1997),
 121-176.

\bibitem{hoppensteadt} F.C. Hoppensteadt \& E.M. Izhikevich, `Weakly Connected Neural Networks',
Springer Verlag (New-York), 1997.

\bibitem{hop}  F.C. Hoppensteadt \& J.P. Keener, {\it{Phase locking of biological clocks}}, J. Math. Biol. {\bf{15}}
(1982), 339-349.

\bibitem{kato} T. Kato, `Perturbation Theory for Linear Operators',
Springer-Verlag (Berlin), 1995.

\bibitem{khadra} A. Khadra \& Y.X. Li, {\it{A model for the pulsatile secretion of gonadotropin-releasing hormone
from synchronized hypothalamic neurons}}, Biophys. J. {\bf{91}} (2006), 74-83.

\bibitem{kruse} K. Kruse \& F. J\"{u}licher, {\it{Oscillations in cell biology}}, Curr. Opin. Cell Bio. {\bf{17}} (2005),
 20-26.

 \bibitem{kuznetsov} A. Kuznetsov, M. K{\ae}rn \& N. Kopell, {\it{Synchrony in a population of
hysteresis-based genetic oscillators}}, SIAM J. Appl. Math. {\bf{65}} (2005) 392-425.


\bibitem{madsen} M.F. Madsen, S. Dan{\o} \& Preben G. S{\o}rensen, {\it{On the mechanisms of glycolytic oscillations in yeast
}}, FEBS Journal, {\bf{272}} (2005),
 2648-2660.

\bibitem{manrubia} S.C. Manrubia, A.S. Mikhailov, D.H. Zanette, `Emergence of Dynamical Order', World Scietific
(New-Jersey), 2004.


\bibitem{pikovsky} A. Pikovsky, M. Rosenblum \& J. Kurths, `Synchronization: a Universal Concept in
Nonlinear Sciences', Cambridge University Press (Cambridge), 2001.


\bibitem{robinson} C. Robinson, `Dynamical Systems', CRC Press (Boca
Raton), 1995.

\bibitem{schibler} U. Schibler \& F. Naef, {\it{Cellular oscillators: rhythmic gene expression and metabolism}},
Current Opinion Cell Biol. {\bf{15}} (2005), 223-229.

\bibitem{toth} R. Toth, A.F. Taylor \& M.R. Tinsley, {\it{Collective behavior of a population of chemically coupled
oscillators}}, J. Phys. Chem. B {\bf{110}} (2006), 10170-10176.

\bibitem{wang} R. Wang \& L. Chen, {\it{Synchronizing genetic oscillators by signalling molecules}},
J. Biol. Rhythms {\bf{20}} (2005), 257-269.

\bibitem{watanabe1} M. Watanabe, {\it{Bifurcation of synchronized periodic solutions in systems
of coupled oscillators I: perturbation results for weak and strong coupling}}, Rocky Mountain J. Math {\bf{23}}
(1993) 1483-1525.
%

\bibitem{winfree} A.T. Winfree, `The Geometry of Biological Time', Springer (New-York), 2001.

\bibitem{wolf} J. Wolf \& R. Heinrich, {\it{Dynamics of two-component biochemical systems in interacting
cells; Synchronization and desynchronization of oscillations and multiple steady states}}, BioSystems {\bf{43}} (1997)
1-24.

\bibitem{wh} J. Wolf \& R. Heinrich, {\it{Effect of cellular interaction on glycolytic oscillations
in yeast: a theoretical investigation}}, Biochem J. {\bf{345}} (2000), 321-334.


\bibitem{zhdanov} V.P. Zhdanov  \& B. Kasemo, {\it{Synchronization of metabolic oscillations: two cells and ensembles of
adsorbed cells}}, J. Biochem. Phys. {\bf{27}} 295-311 (2001).



\end{thebibliography}
\end{document}